\documentstyle[preprint,aps]{revtex}
\tightenlines
\def\dalem{\hbox{$ {  \lower.80ex\hbox{$\sqcap$}
 \atop \raise.80ex\hbox{$\sqcup$}
 }    $}  }
\begin{document}
\preprint{\vbox{Submitted to Phys.\ Rev.\ Lett.\ \hfill}}
                                       
\title{Temperature dependence of instantons in QCD}
\author{M.-C.~Chu$^{(1)}$, S.~M.~Ouellette$^{(2)}$, 
S.~Schramm$^{(3)}$, and R.~Seki$^{(4,5)}$ }
\address{$^{(1)}$ Department of Physics, The Chinese University of Hong Kong,
Shatin, N.T., Hong Kong}
\address{$^{(2)}$ Lauritsen Laboratory, Caltech 452-48, Pasadena, 
CA 91125, U.S.A.}
\address{$^{(3)}$ GSI, Darmstadt, Germany}
\address{$^{(4)}$ W.~K.~Kellogg Radiation Laboratory, Caltech 106-38, 
Pasadena, CA 91125, U.S.A.}
\address{$^{(5)}$ Department of Physics, California State University,
Northridge, CA 91330, U.S.A.}
\date{\today}
\maketitle
 
\begin{abstract}
We investigate the temperature dependence of the instanton contents
of gluon fields, using unquenched lattice QCD and the cooling method.
The instanton size parameter deduced from the correlation 
function decreases from 0.44fm below the phase-transition temperature $T_c$ 
($\approx 150$MeV) to 0.33fm at 1.3 $T_c$.  
The instanton charge distribution is Poissonian above $T_c$, 
but it deviates from the convoluted Poisson at low temperature.  
The topological susceptibility decreases rapidly below $T_c$, 
showing the apparent restoration of the $U(1)_A$ symmetry 
already at $T \approx T_c$.

\end{abstract}
\pacs{PACS numbers: 12.38.Gc, 11.10.Wx, 12.38.Mh}

Among the four global symmetries associated with quantum chromodynamics
(QCD), two chiral symmetries, $SU(N_f)_A$ and $U(1)_A$, must be broken  
as the observed hadron spectrum suggests.  At high temperature, it is
expected that the two symmetries are restored.  
The question as to how they are
restored is an important issue, especially in relation to the QCD
deconfinement phase transition (or maybe a crossover), and is recently
attracting much attention\cite{shur,schafer,cb,two}.  
The deconfinement phase transition 
is thought to have occurred in the evolution of the early universe and 
is expected to generate a new phase of matter, the quark-gluon plasma
(QGP), which is under investigation in relativistic heavy-ion collisions.

Instantons, tunneling between topologically different gauge vacua,
are believed to play crucial roles in the breaking and restoration of
the chiral symmetries, and perhaps also in the deconfinement phase
transition\cite{schafer}:   The $SU(N_f)_A$ symmetry manifests itself
in the Goldstone mode through chiral condensates generated dynamically 
by instantons.  A similar mechanism may be at work in the formation of
hadronic bound states, and the $SU(N_f)_A$ restoration and the
deconfinement are thus closely related to each other.
The $U(1)_A$ symmetry is spontaneously broken by the chiral anomaly
driven by instantons.  At high temperatures, the restoration of the
symmetries takes place because instanton amplitudes are suppressed
through Debye screening, which induces the gluons to be (electrically) 
massive.

How realistic is this instanton mechanism, and what is the precise dynamics
of the instanton participation?  The best means to answer the questions is the 
lattice gauge theory \cite{lat}, which provides a systematic, nonperturbative 
framework to study the equilibrium properties of QCD 
at both zero and finite temperatures.  Whereas direct comparison between
lattice results and experiments is not always possible, lattice QCD often
motivates and constrains phenomenological models, which connect the QCD 
with experiments.  Here, the use of {\it unquenched} QCD must be essential, 
since the instanton mechanism is expected to be affected by the presence of
the dynamical quarks.

In this letter, we report unquenched lattice QCD results on the
temperature dependence of the instanton contents of gluons.
For the first time, we find curious temperature dependence suggesting that 
instantons are seemingly correlated below the $SU(N_f)_A$ phase transition,
$T_c \approx 150$MeV.  The nature and the details of the correlation are
yet unclear, however, and so is the question of whether it corresponds 
to any of the dynamical models\cite{schafer}.   We also find that 
the $U(1)_A$ symmetry appears to be largely already restored
at $T_c$.  Note that because of the cooling method we used, we 
observed neither small instantons nor instanton molecules.

The calculation is carried out for $16^3 \times N_t$ lattices with
$N_t = 4,6,8,10,12,14$, and 16 for $m_q a=0.0125$.  The temperature is varied
by changing $N_t$, with $\beta = 5.54$ fixed.  
The susceptibility and the topological-charge correlation are observed to be  
stable with respect to the number of cooling steps.  
We find that the topological susceptibility decreases slowly at
low temperature as the temperature increases, but a rapid suppression 
sets in around $T = T_c$, so that, compared to the value at $T = 0$,  it is 
already down by an order of magnitude at $T_c$ and becomes only about 1\% 
at $1.3T_c$.  Accompanying this rapid change in $\chi_t$ is
a transition of the instanton size from $\rho \approx 0.44$fm at low
temperature to 0.33fm at $1.3T_c$.  The distribution of instanton charges
is Poissonian above $T_c$, but it deviates from the convoluted 
Poisson at low temperatures.

The instanton content of the gauge fields is monitored by the
the topological-charge density,
which can be measured on the lattice with
\begin{equation}
Q(x_n) = -{1\over 32 \pi ^2} \epsilon _{\alpha \beta \gamma \delta}
{\rm Re \ Tr} \left[ U_{\alpha \beta} (x_n) U_{\gamma \delta} (x_n) \right] \ ,
\label{qn}
\end{equation}
where $U_{\alpha \beta}$ is the product of the link variables around a 
plaquette in the $\alpha - \beta$ plane.  A direct calculation of $Q$
from gauge configurations is plagued with poor statistics and 
with lattice artifacts
associated with discretization.  Teper {\it et al.}~\cite{tep} have shown
that the topological charge can be reliably extracted using the cooling
method, in which one applies a certain number of smoothing steps to 
each configuration in order to minimize the action locally.  The ultraviolet
fluctuations are quickly suppressed by these cooling steps, and only
the relatively stable instantons are left in the configuration, making it
easy to extract the topological charge. Details of the cooling method
are discussed in \cite{tep,chn}.

The topological susceptibility is then given as the fluctuations of the
topological charge:
\begin{equation}
\chi _t \equiv {1 \over N_t N_s ^3 a^4} \langle 
\left( \sum _n Q(x_n) \right) ^2 
\rangle \ \ ,
\label{chi}
\end{equation}
where $\langle ... \rangle$ indicates configuration averaging, $N_t$ 
and $N_s$ are the number of sites in the temporal and spatial direction,
respectively, and $a$ is the lattice spacing.

In order to examine the details of the topological-charge distribution,
we also calculate the topological charge-density correlation function 
\cite{chn},
\begin{equation}
C_Q(x) = \langle \sum _y Q(y) Q(x+y) \rangle / \langle \sum _y Q^2(y) \rangle
\ \ .
\label{cprofile}
\end{equation}
One can compare this correlation function to a convolution of an isolated
instanton topological-charge density
\begin{equation}
Q_\rho (x) = {6 \over \pi ^2 \rho ^4 } \left( {\rho ^2 \over x^2 + \rho ^2 } 
\right) ^4 \ \ . 
\label{instc}
\end{equation}
Zero-temperature calculations using the cooling method \cite{chn} show
that after about 50 cooling steps, the gauge fields are dominated by
large, isolated instantons, whose profiles agree well with Eq. (\ref{instc}).

To generate the unquenched gauge configurations, we adopted and modified 
a code written by the MILC collaboration~\cite{gott}, using its option of 
dynamical Kogut-Susskind fermions.  
Forty configurations are generated for each of $N_t = 12, \ 14,
\ 16$, and 100 configurations for each of $N_t = 4, \ 6, \ 8, \ 10$.
In each case, a molecular-dynamics time step of
$dt=0.02$ is used, and the first 300 time units (43 for $N_t = 12,\ 14, \ 16$)
from a hot start are used as thermalization.  The configurations were
separated by 3 time units for $N_t = 12, \ 14, \ 16$; 5 time units
for $N_t = 4, \ 6, \ 10$; and 10 time units for $N_t = 8$.

These configurations are then cooled, using the standard procedure
as used in \cite{chn}, for 200 steps, saving every 20 steps.  We have
checked to see that both the topological susceptibility and the 
topological-charge correlation function 
settle to stable values by 100 cooling steps.  
The results reported here are based on those extracted after 120 cooling
steps.

The temperature of a gauge configuration is given by 
$T = 1/(N_t a)$ and can be varied by changing either $N_t$, or $a$ ( $ = $ the
lattice spacing).  $a$ is a function of the lattice inverse coupling $\beta$, 
and can be calculated by the perturbation theory, but only for a narrow range 
of $\beta$ (and hence to $T$) in the asymptotic scaling regime.  
In order to obtain  the temperature dependence reliably, we choose to vary 
the temperature by varying $N_t$, keeping $a$ and $\beta$ fixed.  
In this way, the uncertainty in $a$ enters only as an overall constant factor, 
which little affects the functional relation of the $\chi _t$ and $T$.  
The price we pay is that the calculation can be done only for a limited number 
of temperatures corresponding to the discrete values of $N_t$.  
This is especially problematic at high temperatures.
The phase transition with the parameters we use occurs at $N_t = 8$ and a temperature of 
$T_c = 150(9)$ MeV \cite{gott}, giving $a \approx 0.17$fm.

Figure 1 shows the topological susceptibility as a function of the 
temperature.  The solid curve in the figure is the PCAC suppression that is 
due to soft pion gas \cite{sv}, which is believed to be relevant for
low temperature $T << T_c$:
\begin{equation}
\chi _t (T) = \chi _t (T=0) \left( 1 + cT^2 / {F_\pi}^2 \right) \ .
\label{shve}
\end{equation}
Equation~\ref{shve} fits our data very well up to $T = T_c$, 
but with $c = -0.36$,
a value that is well outside the allowed range based on the PCAC argument
$ -1/6 \leq c \leq 1/6$ \cite{sv}.
At higher temperatures, the suppression of the susceptibility
follows a pattern different from the PCAC prediction.  

Also shown in Fig.~1   
is the perturbative Debye-screening prediction (for $N_c = 3, \ N_f = 2$) 
\cite{py}:
\begin{equation}
\chi_t (T) = \chi _t (T=0) \left( 1 + \lambda ^2 /3 \right)^{7 \over 6}
             \exp \left[ -{8 \lambda ^2 /3} -14 \alpha \left( 1 + \gamma
             \lambda ^ {-{3 \over 2}} \right) ^{-8} \right] \ \ ,
\label{piy}
\end{equation}
where $\lambda \equiv \pi \rho T$, $\alpha = 0.01289764$, and 
$\gamma = 0.15858$.  The dot-dashed and dashed curves correspond to 
$\rho = 0.44$ and 0.33 fm, respectively.  Equation~\ref{piy} is not expected
to be valid below $T_c$, but should become relevant at high temperatures.
We observe that the curve for $\rho = 0.44$ fm fits the four lowest
temperature points quite well, giving an alternative
suppression model to PCAC at low temperatures, and that the curve
for $\rho = 0.33$ fm fits the high temperature points quite well. 
This observation suggests that the Debye-screening mechanism may be already 
operative for $T < T_c$, though the dominant instanton size parameter
changes suddenly across $T_c$.  Unfortunately, our $N_t = 4$ ($T=2T_c$) result
is rather uncertain because of severe finite-size effects and because not
one single instanton is found in our 100 configurations.  Also, our method of
temperature variation by varying $N_t$ provides only sparse data 
in the high-temperature regime.  The temperature dependence of the 
Debye-screening predictions must be explicitly verified by 
further calculations.

The general features of the temperature dependence of the $\chi_t$ are the 
same in the quenched calculation\cite{cs} and in the present 
unquenched calculation, but detailed features differ significantly. 
The unquenched $\chi_t$ decreases more slowly at low temperature, but goes 
down much more rapidly around $T_c$.  
The $\chi_t$ is suppressed by almost 90 \% at $T=T_c$ in the unquenched
case, while it is only 50 \% in the quenched case.  That is, 
the suppression of
the instanton amplitudes is enhanced by the dynamical quarks.
The rapid decrease in $\chi_t$ continues above $T_c$, so that only about
1\% remains at $1.3T_c$ in the unquenched, and about 10\% in the quenched 
calculation.
This temperature dependence of $\chi_t$ suggests that the window 
of temperatures  
above $T_c$ in which $U_A(1)$ is broken is {\it narrowed} by the presence 
of dynamical quarks, 
and that $U_A(1)$ is practically restored at about $T_c$. 
One must take caution, however, in applying lattice results on the $U(1)_A$ 
symmetry, because the symmetry is explicitly broken in the conventional lattice 
and the process of approaching the continuum limit involves uncontrolled 
subtlety\cite{cb}.  On the basis of the present calculation with $m_q a=0.0125$, 
we draw no conclusion on the relationship between $T_c$ and 
the $U(1)_A$ symmetry breaking temperature, and thus on whether the phase 
transition is of the first or second order\cite{wilpis,wr}.

Figure 2 shows the temperature dependence of the topological charge correlation 
functions in comparison to the profile of an isolated instanton, Eq.(\ref{instc}). 
In the quenched calculations, $C_Q(r)$ is practically independent of 
the temperature, following remarkably closely to the profile, though hints of a 
transition to a smaller size parameter have been observed above $T_c$\cite{cs}.
The present unquenched calculation confirms the same trend below $T_c$, but  
it clearly shows a rapid transition to a smaller size around $T_c$.
For clarification, we show in Fig.~2 the temperature dependence 
only around $T_c$: the size parameter, $\rho$, is 
approximately 0.44 fm at $0.75 T_c$, shrinks rapidly around $T_c$, and 
becomes about 0.33fm at $1.3T_c$.
It is interesting to note that the low-temperature value of $\rho$ agrees
fairly well with that used in the instanton liquid model,
while the high-temperature value agrees with the dilute-gas model and
also with the quenched value\cite{schafer}.

The distribution of instanton charges in our configuration samples is
shown as histograms for each $N_t$ in Fig.~3.  Whereas in the quenched
case, these distributions agree with the convolued Poisson 
distribution\cite{cs}, 
the unquenched results show significant deviations from this Poissonian,
with curious suppression of the distribution around the zero topological 
charge ($Q =0$) for all $N_t$ below the phase transition.  
The dynamical quarks have apparently
induced non-trivial correlations among the instantons at low temperatures.
Above the phase transition, the distributions become Poissonian, suggesting
a return to the dilute-gas ensemble.

The instantons we have discussed so far in this letter are all of the 
topological charge of the unit magnitude.  We note that we have observed 
occassionally the instantons of the topological charge of the magnitude 
two, which are rarely treated in any dynamical model.
  
In conclusion, our investigation of the temperature dependence of the
instanton contents of unquenched gluon fields yields the following results:
The topological susceptibility decreases rapidly around the 
phase-transition temperature $T_c \approx 150$MeV; at $T = T_c$, it is 
suppressed by one order of magnitude compared to its zero-temperature
value, and at $T = 1.3 T_c$, it is reduced by another order of magnitude. 
The temperature dependence of the topological
susceptibility is consistent with a PCAC prediction for temperatures up
to $T_c$, though with a large coefficient $c$.  It  
also agrees with a Debye-screening model, but with a sudden shrinking of
the instanton size parameter at around $T_c$.
The topological charge correlation function
also changes rapidly around $T_c$, which can be characterized by a
transition of the dominant instanton size parameter from 0.44fm 
at low temperatures to 0.33fm at 1.3 $T_c$. The distribution of the 
instanton charges suggests the existence of nontrivial
correlations among the instantons at low temperatures, which
disappear above $T_c$.   
Our results indicate that the $U_A$(1) symmetry is apparently restored 
largely at $T \approx T_c$. 

This work is in part based on the MILC collaboration's public lattice gauge 
theory code.  We thank C.~DeTar and C.~McNeile for their technical assistance.
We acknowledge San Diego Supercomputer Center for providing Intel Paragon and 
Cray T3E computer resources. 
This research is partially supported by the U.~S.~National Science Foundation
under grants PHY88-17296 and PHY90-13248 at Caltech, the 
U.S.~Department of Energy under grant DE-FG03-87ER40347 at CSUN, 
and the Institute of Mathematical
Sciences at the Chinese University of Hong Kong.  M.~C.~acknowledges the
support of a summer research grant and a direct grant (2060105) of the 
Chinese University of Hong Kong.

\begin{figure}
\caption{
Topological susceptibility as a function of temperature. The data points
correspond to our unquenched-QCD-plus-cooling results with $N_t = 6,  8,  
10,  12,  14,  16$, and $a = 0.17$fm is assumed.  The transition temperature 
$T_c$ is approximately 150 MeV for our calculations with $\beta = 5.54,  
ma= 0.0125, N_x = 16$.
The dashed curve is the prediction of a PCAC model, Eq.~(\ref{shve}), 
with $c = -0.36$; and   
the solid and dot-dashed curves of a Debye-screening model, Eq.~(\ref{piy}), 
with the instanton size parameter, $\rho = 0.44$ and 0.33 fm, respectively. 
}
\label{fig1}
\end{figure}
\begin{figure}
\caption{
Topological charge correlation functions across the phase-transion temperature 
($T_c = 150$MeV) from the same calculations as in Fig.~1.
}
\label{fig2}
\end{figure}

\begin{figure}
\caption{
Histograms showing the distribution of topological charge in the 
configurations for $N_t=6, 8, 10, 12, 14$ and 16. 
$Q$ is the magnitude of the number of topological charges in a configuration, 
and $N(Q)$ is the frequency.  The dots in the figure are the convoluted 
Poisson distributions fitted to the histograms. 
}
\label{fig3}
\end{figure}


\begin{thebibliography}{99}

\let\ul=\underbar
\def\AP#1{{\it Ann.~Phys.~ }{ #1}}
\def\PRP#1{{\it Phys.~Rep.~ }{ #1}}
\def\APP#1{{\it Act.~Phys.~Pol.~ }{ #1}}
\def\PTP#1{{\it Prog.~Theor.~Phys.~ }{ #1}}
\def\PRD#1{{\it Phys.~Rev.~D}{ #1}}
\def\PRC#1{{\it Phys.~Rev.~C}{ #1}}
\def\PRL#1{{\it Phys.~Rev.~Lett.~}{#1}}
\def\PL#1{{\it Phys.~Lett.~ }{ #1}}
\def\NP#1{{\it Nucl.~Phys.~ }{ #1}}
\def\ZP#1{{\it Z.~Phys.~ }{ #1}}
 
\bibitem{shur} E.~Shuryak, Comments Nucl.~Part.~Phys.~{\bf 21}, 235 (1994).

\bibitem{schafer}T.~Sch\"afer and E.~V.~Shuryak, University of Washington 
preprint, hep-ph/9610451, 1997; to appear in Rev. Mod. Phys.

\bibitem{cb} C. Bernard {\it et al.}, Phys. Rev. Lett. {\bf 78}, 598 (1997).

\bibitem{two} G. Boyd, F. Karsch, E. Laermann, and M. Oevers, 
IFUP-TH 40/96 and BI-TP 96/27 (hep-lat/9607046) 1996;
P. deForcrand, M. G. P\'{e}rez, and I-O Stamatescu, hep-lat/9701012, 1997:
and P. deForcrand, M. G. P\'{e}rez, J. E. Hetrick, and I-O Stamatescu, 
hep-lat/9710001, 1997.

\bibitem{lat} I.~Montvay and G.~M\"{u}nster, {\it Quantum Fields on Lattice} 
(Cambridge Univ. Press, Cambridge, 1994), and references therein.  

\bibitem{tep}M.~Teper, Phys.~Lett.~{\bf B171}, 81 (1986); and 
J. Hoek, M.~Teper, and J.~Waterhouse, Nucl.~Phys.~{\bf B288},
 589 (1987).

\bibitem{chn} M.-C.~Chu, J.~M.~Grandy, S.~Huang, and J.~W.~Negele, Phys.~Rev.~
{\bf D49}, 6039 (1994); Phys.~Rev.~Lett.~{\bf 70}, 225 (1993); M.-C.~Chu and 
S.~Huang, Phys.~Rev.~{\bf D45}, 2446 (1992).

\bibitem{cs}M.~-C.~Chu and S.~W.~Schramm, Phys.~Rev.~{\bf D51}, 4580 (1995).

\bibitem{gott} S.~Gottlieb {\it et al.}, Phys.~Rev.~{\bf D47}, 3619 (1993). \\
See http://physics.indiana.edu/\~sg/milc.html. 

\bibitem{sv} E.~Shuryak and M.~Velkovsky, Phys.~Rev.~{\bf D50}, 3323 (1994).

\bibitem{py}R.~D.~Pisarski and L.~G.~Yaffe, Phys.~Lett.~{\bf 97B}, 110 (1980).

\bibitem{wilpis} R.~D.~Pisarski and F.~Wilczek, Phys.~Rev.~{\bf D29}, 338 
(1984).

\bibitem{wr} F. Wilczek, Int. J. Mod. {\bf D3} Suppl., 63 (1994); and
K. Rajogopal and F. Wilczek, Nucl. Phys. {\bf B399}, 395 (1993).

\end{thebibliography}
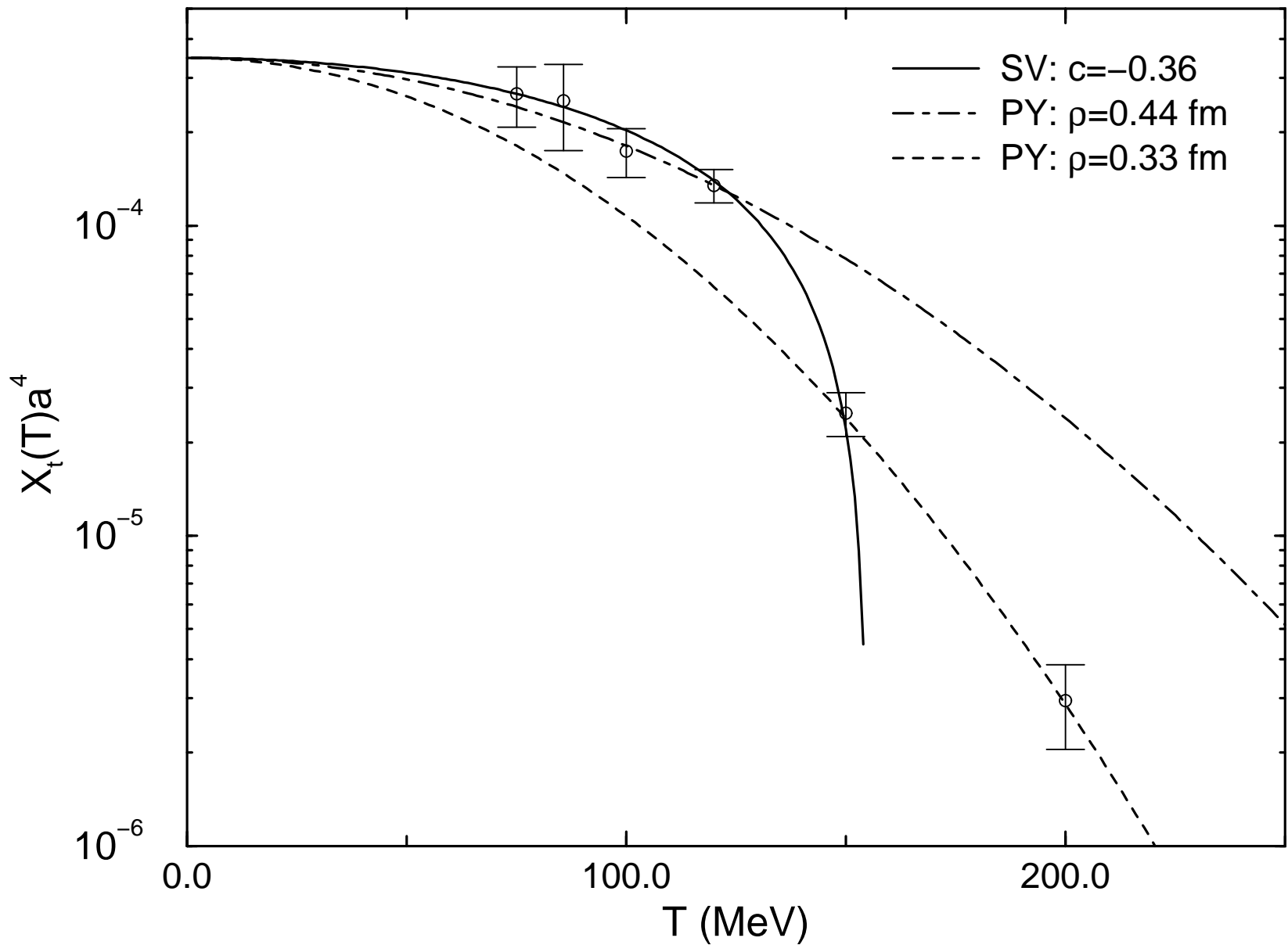
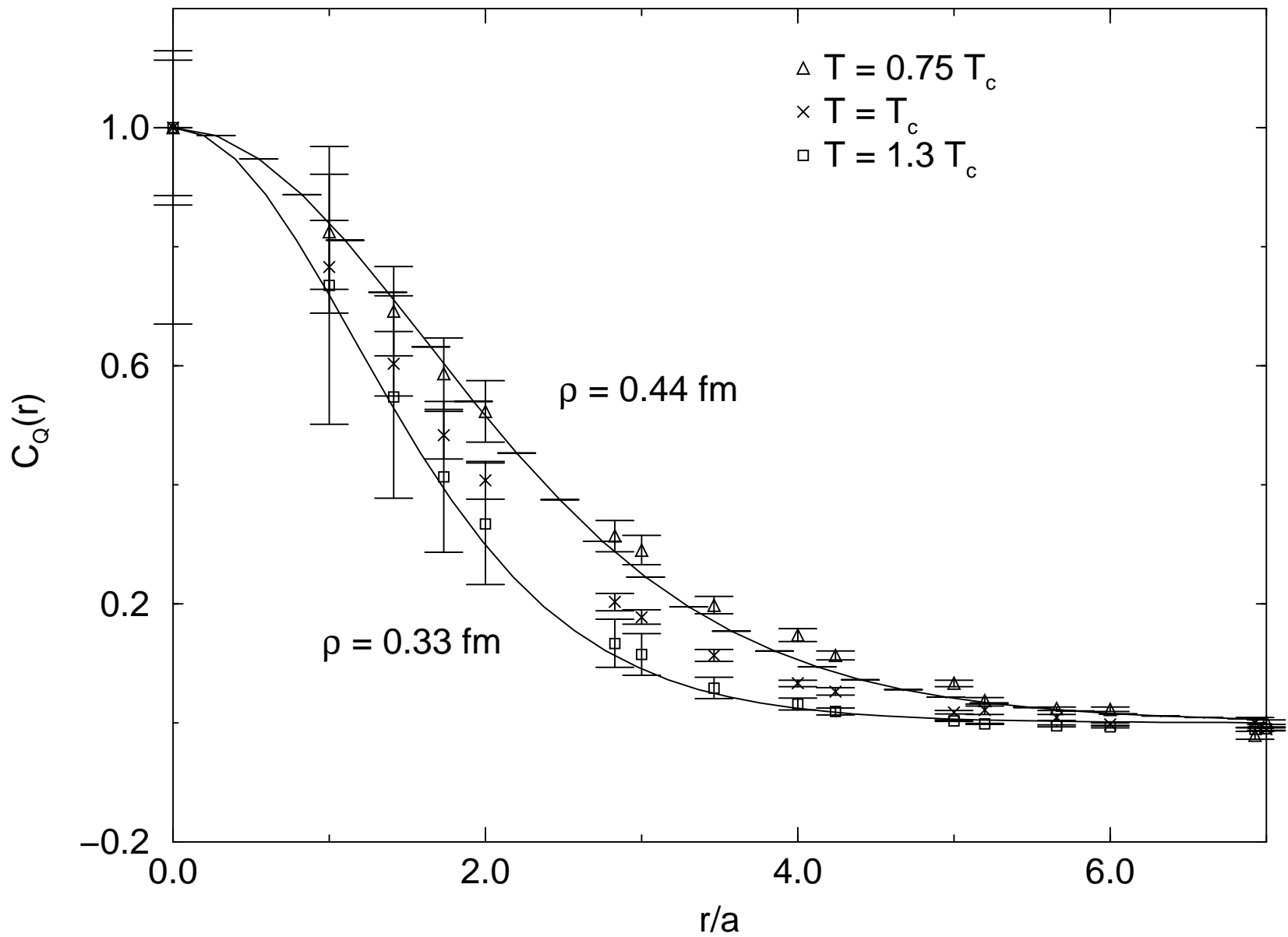
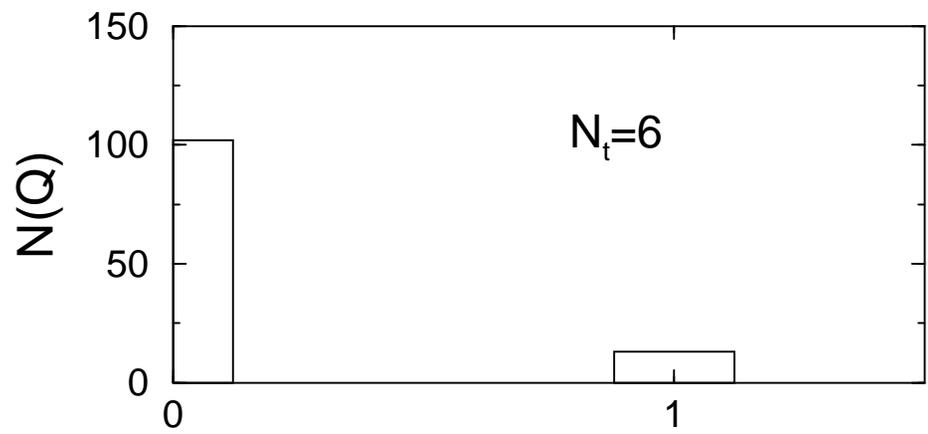
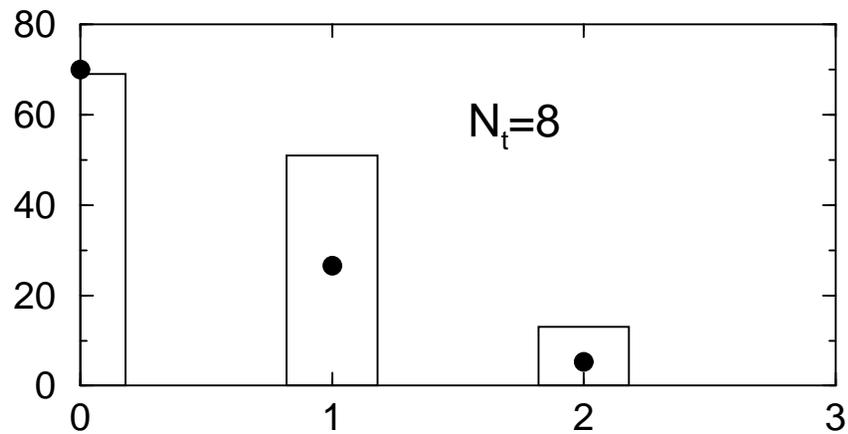
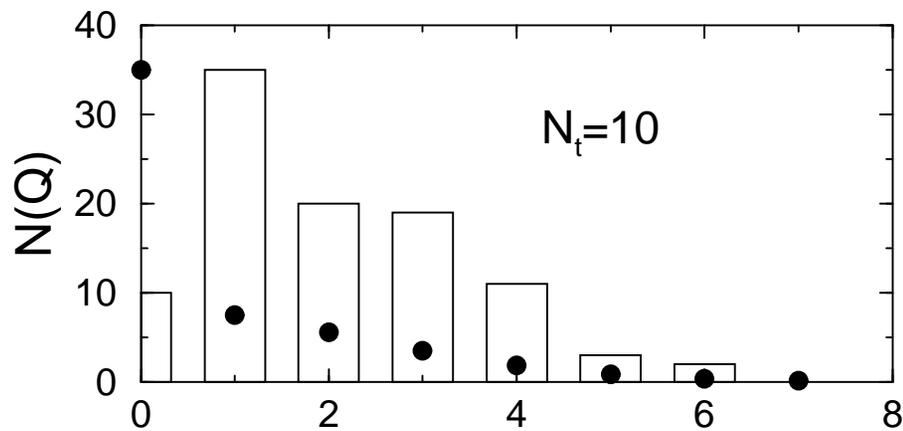
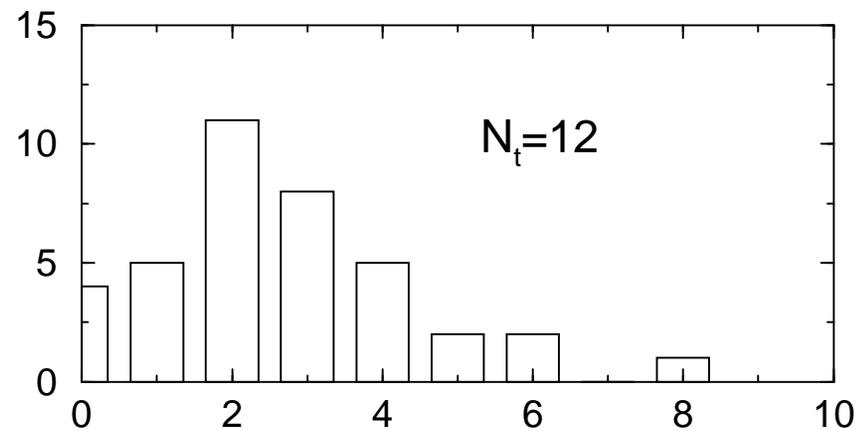
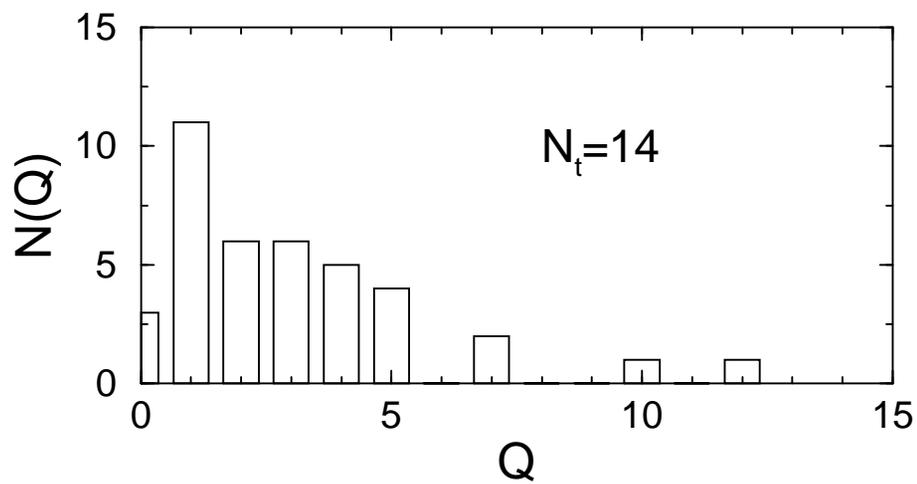
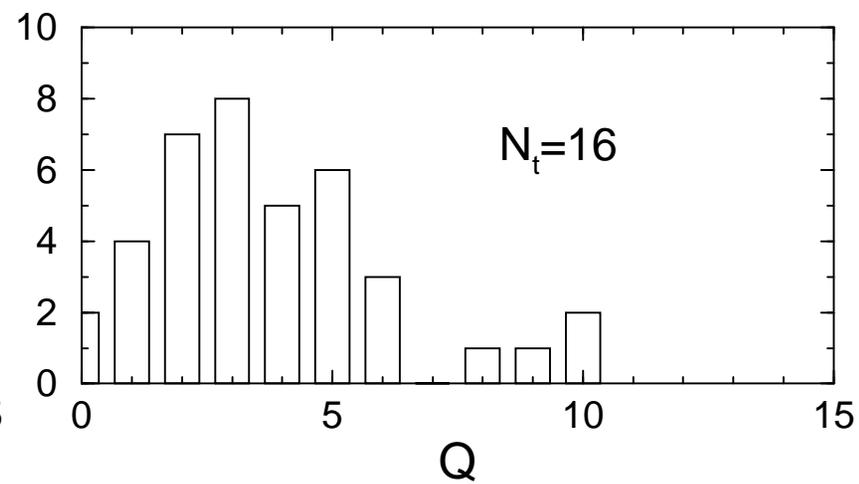
\end{document}